\documentclass{article}
\usepackage{latexsym}
\usepackage{color}
\usepackage{amsmath}
\usepackage{amsfonts}
\usepackage{graphicx}

\title{\textbf{A refined Bogoliubov-Huang approach to helium II thermodynamics}}
\author{Jan Ma\'{c}kowiak and Dawid Borycki\\
Instytut Fizyki, Uniwersytet M. Kopernika w Toruniu,\\
87-100 Toru\'{n}, Poland\\
ferm92@fizyka.umk.pl}

\begin{document}

\maketitle

\begin{abstract}
The thermodynamics of a free Bose gas with effective temperature scale $\tilde{T}$ and hard-sphere Bose gas with the $\tilde{T}$ scale are studied. $\tilde{T}$ arises as the temperature experienced by a single particle in a quantum gas with 2-body harmonic oscillator interaction $V_\textrm{osc}$, which at low temperatures is expected to simulate, almost correctly, the attractive part of the interatomic potential $V_\textrm{He}$ between $^{4}\textrm{He}$ atoms. The repulsive part of $V_\textrm{He}$ is simulated by a  hard-sphere (HS) potential. The thermodynamics of this system of HS bosons, with the $\tilde{T}$ temperature scale (HSET), is investigated, first, by the Bogoliubov-Huang method and next by a modified version of this method, which takes approximate account of those terms of the 2-body repulsion which are linear in the zero-momentum Bose operators $a_0,\,\,a^*_0$ (originally rejected by Bogoliubov). Theoretical heat capacity $C_V(T)$ exhibits good agreement, below 2.1 K, with the experimental heat capacity graph observed in $^{4}\textrm{He}$ at saturated vapour pressure. The phase transition to the low-temperature phase , with a Bose-Einstein condensate, occurs in the HSET at $T_\lambda=$2.17 K, and is accompanied, in the modified HSET version, by a singularity of $C_V(T)$. Other thermal properties of HSET, such as the momentum distribution function, the fraction of atoms in the momentum condensate and normal fluid density, agree qualitatively with those of $^{4}\textrm{He}$, but improve those of the free Bose gas.

\end{abstract}

\section{Introduction}

The extraordinary low-temperature properties of $^{4}\textrm{He}$ have so far evaded full explanation by a first-principles microscopic theory, although the fundamental concepts of such theory are well established. Neutron scattering experiments from liquid helium \cite{PS91} have confirmed London's hypothesis \cite{FL38}, which points to Bose-Einstein condensation (BEC) of $^{4}\textrm{He}$ atoms below $T_\lambda=2.17\,\textrm{K}$, as the decisive factor behind these properties. Theory shows that a free gas of zero-spin bosons (FBG), with particle mass and density equal to those of liquid $^{4}\textrm{He}$, exhibits the BEC transition at $T_0=3.1486\,\textrm{K}$. The shift $T_0-T_\lambda>0$ is therefore presumably a consequence of interatomic interactions.

A decisive step in the understanding of BEC of an interacting gas was made by Bogoliubov \cite{NB47}, \cite{NB70}. His ideas were incorporated into the theory of hard-sphere Bose gas (HSBG) by Lee et al. \cite{LHY57} - \cite{KH95} and were refined by Angelescu et al. \cite{AVZ92}. Unfortunately, the temperature $T_\textrm{BE}$ at which an interacting Bose gas, with $^{4}\textrm{He}$ parameters, exhibits BEC, exceeds $T_0$ (Ref. \cite{KH64}) or remains equal to $T_0$ (Ref. \cite{AVZ92}) in these theories. The HSBG heat capacity in the vicinity of $T=0$ is proportionate to $T^3$, as in $^{4}\textrm{He}$, \cite{KH64} and all atoms belong to the momentum condensate at $T=0\,\textrm{K}$. The experimentally estimated fraction of $^{4}\textrm{He}$ atoms in the momentum condensate (FMC) and fraction of $^{4}\textrm{He}$ normal fluid (FNF) at $T=0\,\textrm{K}$ are equal $0.07 - 0.09$ \cite{GAS2000} and zero, respectively \cite{AL93}.

A remarkable success was achieved by Ceperley and Pollock \cite{CP86} - \cite{DMC95} who computed the canonical density matrix for 64 and 125 interacting $^{4}\textrm{He}$ atoms, using a Monte Carlo path-integral technique. The resulting heat capacity and FNF are in good quantitative agreement with experiment.

 Our objective is to develop a relatively simple theory which would explain, at least some of the unusual properties of liquid $^{4}\textrm{He}$, with accuracy comparable to that achieved in Refs. \cite{CP86} - \cite{DMC95}. This is done by resorting to the concepts of HSBG theory (Refs. \cite{LHY57} - \cite{KH95}), which takes into account the repulsive hard-core part of the 2-body interaction $V_\textrm{He}$ between $^{4}\textrm{He}$ atoms. The attractive part of $V_\textrm{He}$ is simulated by a harmonic oscillator 2-body potential $V_\textrm{osc}$, which gives rise to an effective temperature $\tilde{T}$ of the gas. $\tilde{T}$ is the effective temperature experienced by a single particle in a quantum Boltzmann gas with 2-body oscillator interactions. It is found in Section 2 by examining the structure of the $1-\textrm{particle}$ reduced density matrix $\tilde\varrho^{(1)}$ of the canonical $N-\textrm{particle}$ density matrix $\varrho^{(N)}$ of such gas with, additionally, a harmonic oscillator attraction $V_0$ between each particle and a fixed centre. For large $N$, one finds that, for sufficiently weak $V_0$, $\tilde\varrho^{(1)}\approx\textrm{const}\exp[-\tilde\beta H_0^{(1)}]$, where $H_0^{(1)}=-\hbar^2\triangle/2m$ and
\begin{equation}
\tilde\beta=(k_\textrm{B}\tilde{T})^{-1}=\gamma^{-1}\tanh(\gamma\beta)
\end{equation}
with
\begin{equation}
\gamma=\frac{\hbar}{2}\sqrt{\frac{Nk}{m}},\qquad \beta=(k_\textrm{B}T)^{-1},
\end{equation}
$k$ denoting the coupling constant of $V_\textrm{osc}$.

The procedure of reducing $\varrho^{(N)}$ amounts (up to a multiplicative constant $|\Lambda|^{-N+1}$, where $|\Lambda|$ denotes the system's volume) to averaging $\varrho^{(N)}$ over the positions of $N-1$ particles of the $N$-particle system. Thus, from the viewpoint of $1$-particle measurements, a quantum Boltzmann gas of particles interacting via $V_\textrm{osc}$ behaves like a gas of free particles with the same mass but at a different temperature $\tilde{T}$. We put forward a conjecture that a Bose gas with the 2-body interaction $V_\textrm{osc}$ behaves analogously. This assumption is the starting point for further investigation.

The next stages of this inquiry consist in deriving the thermodynamics of a free Bose gas with the temperature scale $\tilde{T}$ (FBET) (Sec. 3) and hard-sphere Bose gas with this temperature scale (HSET) according to the Bogoliubov-Huang approach (Sec. 4) . It is shown that the HSET allows to improve the FBET description of low-temperature thermodynamics of $^{4}\textrm{He}$. In particular, the HSET heat capacity shows quantitative agreement with the low-temperature branch of experimental $^{4}\textrm{He}$ heat capacity below 1.8 K. Other HSET thermal properties agree qualitatively with those of $^{4}\textrm{He}$, e.g., FMC and FNF, at $T=0\,\textrm{K}$, equal 0.5224 and 0.1896, respectively.

An alternative version of the HSET, which takes approximate account of those terms of the 2-body repulsion, which are linear in the zero-momentum Bose operators $a_0,\,\,a^*_0$ (originally rejected by Bogoliubov in Ref. \cite{NB47}) is studied in Sec. 5. This modified HSET (DHSET) yields singularity of heat capacity at $T_\lambda$ and $C_V(T)$ agrees with experiment below 2.1 K. The FMC and FNF in this approach equal 0.4266 and 0.1178, respectively, at $T=$ 0 K. The DHSET thus considerably improves the HSET description of helium II thermodynamics.

\section{The effective temperature}

The interatomic potential $V_\textrm{He}$ has a typical well with a minimum at the distance $r_0=3{\AA}$ separating two $^{4}\textrm{He}$ atoms \cite{RA79}. In the low-temperature regime, a reliable approximation to the partition function
\begin{equation}
Z= \textrm{Tr}(\exp[-\beta H^{(N)}]S^{(N)}),
\end{equation}
where
\begin{equation}
H^{(N)}=\sum_{i=1}^N H_0(\textbf{r}_i)+\sum_{1\leq i<j\leq N}V_\textrm{He}(|\textbf{r}_i-\textbf{r}_j|)
\end{equation}
and $S^{(N)}$ denotes the symmetrizer, can be therefore obtained by applying the Laplace method, i.e., by expanding $V_\textrm{He}(r)$ in the neighbourhood of $r_0$ up to second order and integrating the resulting Gaussian function. This procedure amounts to replacing $V_\textrm{He}(|\textbf{r}_1-\textbf{r}_2|)$, up to an additive constant, by
\begin{equation}
V_2(|\textbf{r}_1-\textbf{r}_2|)=\frac{k}{2}(|\textbf{r}_1-\textbf{r}_2|-r_0)^2,\quad k>0.
\end{equation}
The approximating partition function
\begin{equation}
Z_2=\textrm{Tr}(\exp[-\beta H_2^{(N)}]S^{(N)}),
\end{equation}
where
\begin{equation}
H_2^{(N)}=\sum_{i=1}^N H_0(\textbf{r}_i)+\sum_{1\leq i<j\leq N}V_2(|\textbf{r}_i-\textbf{r}_j|),
\end{equation}
is intractable, unless we put $r_0=0$. We are then faced with the problem of solving the thermodynamics of $N$ bosons coupled by harmonic oscillator interactions. The eigenproblem for the Hamiltonian
\begin{equation}
H_\textrm{osc}^{(N)}=H_2^{(N)}|_{r_0=0}
\end{equation}
of such system, with the Hilbert space $S^{(N)}L^2(\mathbb{R}^3)^N$, was solved in Ref. \cite{MZ00}. The grand partition function $Z_\textrm{osc}$ for $H_2^{(N)}$, with $r_0=0$, expresses in closed form in terms of the fugacity $z$. However, the equation for $z$ takes the same form as for uncoupled oscillators ( Ref. \cite{MZ00}), viz.,
\begin{equation}
N=\sum_{n_x,n_y,n_z=0}^\infty(z^{-1}\exp[\beta\hbar\omega(n_x+n_y+n_z)]-1)^{-1},
\end{equation}
and is also intractable, unless approximations are applied (e.g. Ref.\cite{FD99}).

The eigenstructure of $S^{(N)}H_2^{(N)}S^{(N)}$, with $r_0=0$, found in Ref. \cite{MZ00}, appears to be inadequate for the description of $^{4}\textrm{He}$ thermodynamics, also from the viewpoint of existing theories of BEC in homogeneous systems, which usually exploit the Hamiltonian $H_{0\Lambda}^{(N)}+V_\Lambda^{(N)}$ of $N$ bosons, with an interaction $V_\Lambda$, enclosed in a cube $\Lambda=L^3$, under periodic boundary conditions. Owing to the different Hilbert spaces of $S^{(N)}H_\textrm{osc}^{(N)}$ and $S^{(N)}H_{0\Lambda}^{(N)}$, the thermodynamics of the first Hamiltonian does not pass over to that of the later in the limit $k\rightarrow 0$, as one would require for a Bose gas with weak $V_2|_{r_0=0}$ and the usual infinite-volume limit: $|\Lambda|\rightarrow\infty, N\rightarrow\infty, N/|\Lambda|=\textrm{const}$, cannot be performed for $S^{(N)}H_\textrm{osc}^{N}$.

A possible approximation to
\begin{equation}
\varrho_\textrm{osc}^{(N)}=Z_\textrm{osc}^{-1}\exp[-\beta H_\textrm{osc}^{(N)}]
\end{equation}
in  $S^{(N)}L^2(\Lambda)^N$, which fulfils this consistency requirement, can be obtained by exploiting the structure of the reduced $1$-particle density matrix of the density operator
\begin{equation}
\varrho_1^{(N)}=Z_1^{-1}\exp[-\beta H_1^{(N)}],\quad Z_1=\textrm{Tr}\exp[-\beta H_1^{(N)}],
\end{equation}
defined in $L^2(\mathbb{R}^3)^N$, where
\begin{equation}
H_1^{(N)}=\sum_{i=0}^N(H_0(\textbf{r}_i)+\frac{k_0}{2}r_i^2)+\frac{k}{2}\sum_{i\leq i<j\leq N}(\textbf{r}_i-\textbf{r}_j)^2.
\end{equation}
The reduced density matrices of $\varrho_1^{(N)}$ were studied in Refs. \cite{SP71}-\cite{SP74}. It was shown in Refs. \cite{SP71}, \cite{SPA72} that the $1$-particle reduced density matrix $\varrho_1^{(1)}=L_N^1\varrho_1^{(N)}$ ($L_N^1$ denoting the partial trace over $L^2(\mathbb{R}^3)^{N-1}$) has the form
\begin{equation}
\varrho_1^{(1)}=\zeta_1^{-1}\exp[-\beta_1h^{(1)}],\quad h^{(1)}=H_0^{(1)}+\frac{k_1}{2}r^2,
\end{equation}
where
\begin{equation}
\zeta_1=\big(2\sinh(\frac{1}{2}\Omega_1)\big)^{-3},\quad \Omega_1=\hbar^2\alpha_1^2\beta_1/m,\quad
\alpha_1^2=\sqrt{mk_1}/\hbar
\end{equation}
and $k,\, k_0,\, k_1,\,\beta,\,\beta_1$ are related by the following formulae :
\begin{equation}
k_1=\frac{k_0k_N^{1/2}\tanh(\frac{1}{2}\Omega_N)+k_Nk_0^{1/2}(N-1)\tanh(\frac{1}{2}\Omega_0)}
{k_N^{1/2}\tanh(\frac{1}{2}\Omega_N)+k_0^{1/2}(N-1)\tanh(\frac{1}{2}\Omega_0)},
\end{equation}
\begin{equation}
\frac{N^2}{\sqrt{k_0k_N}}\coth\Big(\frac{\hbar\beta_1}{2}\sqrt{\frac{k_1}{m}}\Big)^2=\nonumber
\end{equation}
\begin{equation}
=\Big(\frac{\coth(\frac{1}{2}\Omega_0)}{\sqrt{k_N}}+(N-1)\frac{\coth(\frac{1}{2}\Omega_N)}{\sqrt{k_0}}\Big)
\Big(\frac{\coth(\frac{1}{2}\Omega_0)}{\sqrt{k_0}}+(N-1)\frac{\coth(\frac{1}{2}\Omega_N)}{\sqrt{k_N}}\Big),
\end{equation}
with
\begin{equation}
\Omega_0=\hbar\beta\sqrt{\frac{k_0}{m}},\quad\Omega_N=\hbar\beta\sqrt{\frac{k_N}{m}},\quad k_N=k_0+Nk.
\end{equation}
In the limit $k_0\rightarrow 0$, $k_1$ approaches zero linearly in $k_0$:
\begin{equation}
k_1=a k_0+\textrm{O}(k_0^2)\quad \textrm{as}\quad k_0\rightarrow 0,
\end{equation}
where
\begin{equation}
a=\frac{\tanh(\gamma\beta)+(N-1)\gamma\beta}{\tanh(\gamma\beta)},\quad \gamma=\frac{\hbar}{2}\sqrt{\frac{Nk}{m}}.
\end{equation}
Using Eqs. (16), (18), one finds
\begin{equation}
\lim_{k_0\rightarrow0}\beta_1(\beta)=\frac{N\beta\tanh(\gamma\beta)}{\tanh(\gamma\beta)+(N-1)\gamma\beta}=
\tilde{\beta}_1(\beta).
\end{equation}
Eqs. (12), (13), (18), (20) show that, if measurements are restricted to $1$-particle observables, a single particle of a quantum Boltzmann gas in the state
\begin{equation}
\varrho_2^{(N)}=\textbf{Z}^{-1}\exp[-\beta H_2^{(N)}|_{r_0=0}]
\end{equation}
behaves like one belonging to a free quantum gas at an effective temperature $\tilde{T}_1=(k_\textrm{B}\tilde{\beta}_1(\beta))^{-1}$.
We expect that for a macroscopic sample of a Bose gas with the Hamiltonian (12), the meaning of $\tilde{\beta}_1$ is analogous. In other words, by resorting to the asymptotic form of the 1-particle reduced density matrix of a free Bose gas \cite{JM99}, we put forward the following conjecture: for sufficiently small $k_0>0$,
\begin{equation}
Z_{1S}^{-1}L_N^1(S^{(N)}\exp[-\beta H_1^{(N)}])\approx\frac{1}{N}z\tilde{\varrho}_1^{(1)}(1-z\tilde{\varrho}_1^{(1)})^{-1},
\end{equation}
where
\begin{equation}
Z_{1S}=\textrm{Tr}(S^{(N)}\exp[-\beta H_1^{(N)}]),\quad\tilde{\varrho}_1^{(1)}=\exp(-\tilde{\beta}_1H_0^{(1)}).\nonumber
\end{equation}
Clearly, if $T=(k_\textrm{B}\beta)^{-1}$ denotes the real temperature, then $\tilde{T}_1>T$, if $\gamma>0$, and
\begin{equation}
\lim_{\gamma\rightarrow0}\tilde{T}_1=T,
\end{equation}
as required by consistency.

It is worth noting that the effective temperature of a quantum gas is not an entirely new concept. For example Landau and Lifshitz in Ref. \cite{LL51} (Sec. 33) computed the effective temperature of an interacting quantum gas, using a semi-classical approximation.

Let us now define
\begin{equation}
\tilde{\beta}(\beta)=\lim_{N\rightarrow\infty}\tilde{\beta}_1(\beta)=\gamma^{-1}\tanh(\gamma\beta).
\end{equation}
This limiting form $\tilde{\beta}$ of $\tilde{\beta}_1$ provides the best agreement with experimental data on $^{4}\textrm{He}$.
Clearly, $\tilde{\beta}\rightarrow\beta$ as $\gamma\rightarrow0$. This property of $\tilde{\beta}$, as well as the construction of $\tilde{\varrho}_1^{(1)}$, allow to expect that a sufficiently weak harmonic oscillator interaction, between particles of a quantum gas in equilibrium, can be simulated by performing the substitution $\beta\rightarrow\tilde{\beta}$. In this approach, the thermodynamics of a gas enclosed in a finite volume $\Lambda$, with the 2-body oscillator interaction accounted for by the temperature scale $\tilde{T}$, passes over to the thermodynamics of a free gas in $\Lambda$, with the real scale $T$, when this interaction is switched off. In this manner, the consistency requirement, discussed above, is fulfilled. The consequences of this approach will be investigated in the next sections.

The same effective temperature $\tilde{\beta}$ of a quantum gas, in a field of randomly located oscillator wells, representing the Coulomb attraction between particles of the gas and localized metallic ions,  was found in Ref. \cite{JM07} by averaging the canonical density matrix of this system over the positions of the wells. A simple solution of the Kondo problem, in terms of $\tilde{T}$, was given in Ref. \cite{JM08}. The well known puzzling dependence of $T_\textrm{c}$ on dopant concentration in high-$T_\textrm{c}$ superconductors, was explained also in terms of $\tilde{T}$ in Ref. \cite{MB10}, as an effect arising from these interactions.

\section{Free Bose gas with the effective temperature scale $\tilde{T}$}

The fugacity of the free Bose gas (FBG) equals \cite{KH63}
\begin{eqnarray}
z=\Bigg\{\begin{array}{ll}
1,&T\leq T_\textrm{c},\quad\\
\textrm{the root of }g_{3/2}(z)=d\lambda^3
,&T\geq T_\textrm{c},
\end{array}
\end{eqnarray}
where the BEC transition temperature $T_\textrm{c}$ equals
\begin{equation}
T_\textrm{c}=\Big(\frac{d}{g_{3/2}(1)}\Big)^{2/3}\frac{2\pi\hbar^2}{k_\textrm{B}m}
\end{equation}
and
\begin{equation}
g_s(x):=\sum_{k=1}^{\infty}\frac{x^k}{k^s},\quad \lambda=\Big(2\pi\hbar^2\beta/m\Big)^{1/2},\quad d=\frac{N}{|\Lambda|}.
\end{equation}
Since $\tilde{T}(T)$ is a monotonously increasing function, the fugacity of the free Bose gas, with effective temperature scale $\tilde{T}$ (FBET), equals
\begin{eqnarray}
\tilde{z}=\Bigg\{\begin{array}{ll}
1,&T\leq \tilde{T}_\textrm{c},\quad\\
\textrm{root of }g_{3/2}(\tilde{z})=d\tilde{\lambda}^3,&T\geq\tilde{T}_\textrm{c},
\end{array}
\end{eqnarray}
with $\tilde{T}_\textrm{c}$ defined by the equality
\begin{equation}
\tilde{\beta}(\tilde{T}_\textrm{c})=\Big(\frac{g_{3/2}(1)}{d}\Big)^{2/3}\frac{m}{2\pi\hbar^2}
\end{equation}
and
\begin{equation}
\tilde{\lambda}=\Big(2\pi\hbar^2\tilde{\beta}/m\Big)^{1/2}.
\end{equation}
The expressions for the pressure, free energy and energy of FBET are analogous to those for the FBG :
\begin{equation}
\tilde{P}_0=\frac{g_{5/2}(\tilde{z})}{\tilde{\beta}\tilde{\lambda}^3},
\end{equation}
\begin{equation}
\tilde{F}_0=-\frac{N}{\tilde{\beta}}\Big(\frac{g_{5/2}(\tilde{z})}{\tilde{d\lambda}^3}+\ln\tilde{z}\Big),
\end{equation}
\begin{equation}
\tilde{U}_0=\frac{3Ng_{5/2}(\tilde{z})}{2\tilde{\beta}d\tilde{\lambda}^3}.
\end{equation}
For the heat capacity $\tilde{C}_0$ and entropy $\tilde{S}_0$ of the BGET one obtains
\begin{eqnarray}
\tilde{C}_0=\Bigg\{\begin{array}{ll}
\frac{15Ng_{5/2}(\tilde{z})}{4k_\textrm{B}\tilde{\beta}^2\tilde{\lambda}^3\cosh^2(\gamma\beta)T^2d}
+\frac{3Ng_{3/2}(\tilde{z})}{2\tilde{\beta}\tilde{\lambda}^3\tilde{z}d}\frac{\partial\tilde{z}}{\partial T},
&T\geq\tilde{T}_\textrm{c},\quad\\
\quad & \quad\\
\frac{15Ng_{5/2}(1)}{4k_\textrm{B}\tilde{\beta}^2\tilde{\lambda}^3\cosh^2(\gamma\beta)T^2d},&T\leq\tilde{T}_\textrm{c},
\end{array}
\end{eqnarray}
where
\begin{equation}
\frac{\partial\tilde{z}}{\partial T}=\frac{3}{2}\tilde{\lambda}^3\tilde{z}d\frac{\partial\tilde{\beta}}{\partial T} \Big(\tilde{\beta}
g_{1/2}(\tilde{z})\Big)^{-1},
\end{equation}
and
\begin{equation}
\tilde{S}_0=-\frac{\partial\tilde{F}}{\partial T}=N\Big(\tilde{\beta}^2\cosh^2(\gamma\beta)k_\textrm{B}T^2\Big)^{-1}\Big\{\frac{5g_{5/2}(\tilde{z})}{2 \tilde{\lambda}^3d}+\ln\tilde{z}\Big\}.
\end{equation}
The FBET, with $m$ equal to the mass of $^{4}\textrm{He}$ atom: $m=m_\textrm{He}=6.64765\,10^{-24}\textrm{g}$ and density of $^{4}\textrm{He}$ at $T_{\lambda}$ under saturated vapour pressure (SVP): $d=d_{\textrm{He}}=0.02197{\AA}^{-3}$ \cite{JW63}, has the advantage that it allows to lower the BEC transition temperature below the FBG value $T_0$. Using Eq. (35), one obtains for $m=m_\textrm{He}$, $d=d_{\textrm{He}}$,
\begin{equation}
\tilde{C}_0=4.93184008\, 10^8\frac{\partial\tilde{\beta}}{\partial T}\Big\{-\frac{15}{4}g_{5/2}(\tilde{z})\tilde{\beta}^{-7/2}+\nonumber
\end{equation}
\begin{equation}
\frac{9}{4}g_{3/2}(\tilde{z})g_{3/2}(1)(k_\textrm{B}T_0)^{3/2}\Big(\tilde{\beta}^2g_{1/2}(\tilde{z})\Big)^{-1}\Big\}
\frac{\textrm{cal}}{\textrm{g K}}
\end{equation}
if $\gamma$ is expressed in eV and $k_\textrm{B}$ in eV/K. The graphs of $\tilde{C}_0$ for various values of $\gamma$ are plotted Fig. 1. Experimental data $C_\textrm{He}(T)$ on heat capacity of $^{4}\textrm{He}$ at SVP, from Refs. \cite{JW63}, \cite{GA73}, are also presented. The plot for $\gamma=0$ represents heat capacity of FBG with $m=m_\textrm{He},\,\,d=d_\textrm{He}$.
\begin{figure}
\scalebox{1.1}{\includegraphics{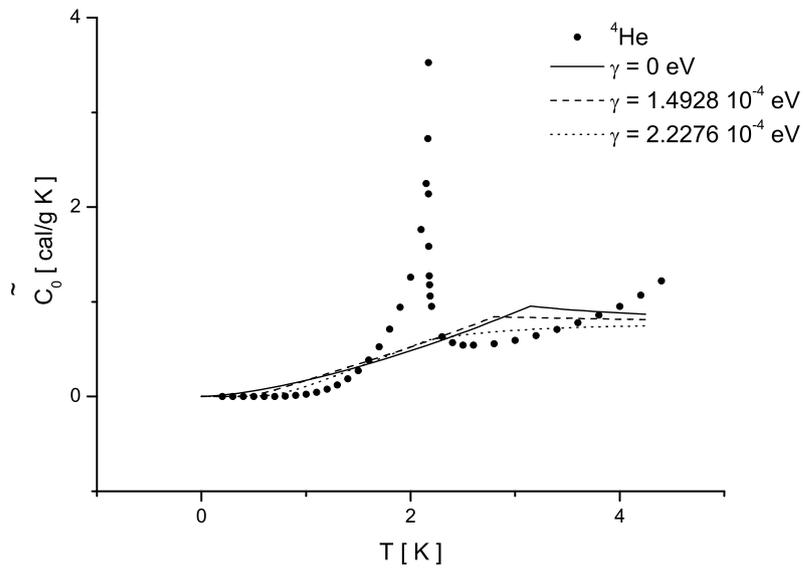}}
\caption{Heat capacity $\tilde{C}_0$ of FBET with $m=m_\textrm{He},\,\,d=d_{\textrm{He}}=0.02197 {\AA}^{-3}$, under varying $\gamma$. $\gamma=0$ corresponds to FBG with absolute temperature scale $T$. The points are experimental results from Refs. \cite{JW63}, \cite{GA73}.}\label{Fi:CVFGET}
\end{figure}

Disagreement between the $\tilde{C}_0(T)$ plots and $C_\textrm{He}(T)$ at $T\leq T_\lambda$ is considerable. However, the $\tilde{C}_0(T)$ plot for $\gamma=1.4928\, 10^{-4}\,\textrm{eV}$, with $\tilde{T}_\textrm{c}=2.8$ K, provides a slight improvement of the $\tilde{C}_0(T)$ plot for $\gamma=0$ (FBG). A further improvement, below $1\,\textrm{K}$, is visible in the $\tilde{C}_0(T)$ plot with $\tilde{T}_\textrm{c}=T_\lambda,\,\,\gamma=\gamma_0=2.2276\,10^{-4}\,\textrm{eV}$, although at $T_\lambda$ there is no peak.

The fraction of atoms in the momentum condensate (FMC) and zero latent heat, accompanying the $\lambda$-transition in $^{4}\textrm{He}$, are also more conveniently described in terms of the FBET than FBG. The FMC for the FBG equals
\begin{equation}
\xi_0(T)=1-\Big(\frac{T}{T_\textrm{c}}\Big)^{3/2}.
\end{equation}
The corresponding quantity for the BGET therefore takes the form
\begin{equation}
\tilde{\xi}_0(T)=1-\Big(\frac{\tilde{\beta}_\textrm{c}}{\tilde{\beta}(T)}\Big)^{3/2},
\end{equation}
where $\tilde{\beta}_\textrm{c}=\tilde{\beta}(\gamma,\tilde{T}_\textrm{c})$. The values of $\tilde{\xi}_0(T=0)$ are given in Table 1. For $\gamma>0$, FMC satisfies $\tilde{\xi}_0(0)<1$, similarly as in $^{4}\textrm{He}$, where FMC is estimated to be equal $0.07-0.09$ \cite{GAS2000}, \cite{AL93}.

There is no latent heat accompanying the $\lambda$-transition in $^{4}\textrm{He}$, whereas
BEC of the FBG is a 1-st order transition and the accompanying latent heat per atom equals \cite{KH63}
\begin{equation}
L_0=\frac{5}{2}\frac{g_{5/2}(1)}{g_{3/2}(1)}k_\textrm{B}T_0.
\end{equation}
Huang derives Eq. (40), using the Clapeyron equation
\begin{equation}
\frac{\partial P}{\partial T}=\frac{L}{T\Delta v}
\end{equation}
and the fact the specific volume of the BE condensate of FBG equals zero, implying
\begin{equation}
\Delta v=v_\textrm{c}=\frac{\lambda^3}{g_{3/2}(1)}.
\end{equation}
Analogously, for the FBET,
\begin{equation}
\Delta v=\tilde{v}_\textrm{c}=\frac{\tilde{\lambda}^3}{g_{3/2}(1)}
\end{equation}
and the vapour pressure at $T<\tilde{T}_\textrm{c}$ equals
\begin{equation}
\tilde{P}_0=\frac{g_{5/2}(1)}{\tilde{\beta}\tilde{\lambda}^3}.
\end{equation}
Hence,
\begin{equation}
\frac{\partial\tilde{P}_0}{\partial T}=\frac{5}{2}\frac{g_{5/2}(1)}{k_\textrm{B}T^2\tilde{\beta}^2\cosh^2(\gamma\beta)
\tilde{v}_\textrm{c}g_{3/2}(1)}=\frac{\tilde{L}_0}{T\tilde{v}_\textrm{c}}.
\end{equation}
The latent heat per atom, at the transition of the BGET, therefore equals
\begin{equation}
\tilde{L}_0=\frac{5}{2}
\frac{g_{5/2}(1)}
{k_\textrm{B}\tilde{T}_\textrm{c}\tilde{\beta}(\tilde{T}_\textrm{c})^2
\cosh^2(\gamma/(k_\textrm{B}\tilde{T}_\textrm{c}))g_{3/2}(1)}.
\end{equation}
Furthermore, $\tilde{L}_0$ is correctly related to the entropy difference $\Delta\tilde{S}_0$ between the gas phase and the momentum condensate. The entropy of the latter equals zero, hence $\Delta\tilde{S}_0(T)=\tilde{S}_0(T)$. Thus,  $\tilde{L}_0=\tilde{T}_\textrm{c}\Delta\tilde{S}_0/N$. The decreasing values of $\tilde{L}_0$, for increasing $\gamma$, are given in Table 1.

Table 1

\begin{equation}
\begin{array}{llll}
\tilde{T}_\textrm{c}\,\,[\textrm{K}]&\quad\gamma\,[\textrm{eV}]&\tilde{\xi}_0(0)&\quad\tilde{L}_0\,[\textrm{eV}]\\
\quad&\quad&\quad&\quad\\
3.1486&\quad0&\quad1&L_0=3.448585\,10^{-4}\\
2.8&1.4928\,10^{-4}&0.59186&\quad2.7306\,10^{-4}\\
2.17&2.2276\,10^{-4}&0.23187&\quad1.4982\,10^{-4}\\
\end{array}
\end{equation}

The above discussion shows that the FBET has several advantages over the FBG, as a model of low temperature $^4{\textrm{He}}$ thermodynamics. However, adjustment of $\tilde{T}_\textrm{c}$ to $T_\lambda$ requires setting $\gamma=\gamma_0$, which leads to a flattened $\tilde{C}_0(T)$ graph without any peak. Further modification of the FBET is therefore necessary.

\section{Bogoliubov-Huang approach to hard-sphere Bose gas with $\tilde{T}$ scale}

The substitution $V_\textrm{He}\rightarrow V_2(r_0=0)$, performed in Section 2, amounts to considering only the attractive part of $V_\textrm{He}$ in a close vicinity of the minimum. Furthermore, the thermal properties of a gas with such attraction are the same as those of a free gas with the temperature scale $\tilde{T}$, the FBET, if measurements are restricted to $1$-particle observables. Discrepancies between these properties and those of $^4{\textrm{He}}$ show that the 2-body interaction considered so far is oversimplified and, therefore, that the repulsive character of $V_\textrm{He}$, at small distances, should be also accounted for by theory. These observations lead naturally to the concept of a hard-sphere Bose gas with the effective temperature scale $\tilde{T}$ (HSET).

The thermodynamics of the hard-sphere Bose gas (HSBG) was studied by Lee et al. \cite{LHY57} - \cite{KH95}. The Hamiltonian has the form
\begin{equation}
H=\sum_\textbf{k}(\varepsilon_\textbf{k}-\mu)a_\textbf{k}^{*}a_\textbf{k}
+\frac{g}{2|\Lambda|}{\sum_\textbf{k}}^{'}\sum_\textbf{p,q}
a_\textbf{p}^{*}a_\textbf{q}^*a_\textbf{p+k}a_\textbf{q-k},\nonumber
\end{equation}
where $\varepsilon_\textbf{k}=\hbar^2k^2/2m,\,\,g=4\pi a\hbar^2/m,\,\,a$ denoting the hard-sphere diameter and
\begin{equation}
{\sum_\textbf{k}}^{'}f(k):=\sum_\textbf{k}[f(k)-\frac{A}{k^2}],
\end{equation}
if $f(k)=A k^{-2}+O(k^{-(3+\eta)}),\,\eta>0$, as $k\rightarrow\infty$.
According to Bogoliubov \cite{NB47}, \cite{NB70}, the interaction terms in $H$, crucial for the description of BEC, are those containing $a_0^2,\,a_0^{*2}$ or $a_0^*a_0$, owing to the observed macroscopic occupation of the zero-momentum 1-particle state.  By cancelling the remaining terms, Bogoliubov obtains the following reduced Hamiltonian :
\begin{equation}
H_\textrm{r}=\sum_\textbf{k} (\varepsilon_k-\mu) a^*_\textbf{k}a_\textbf{k}+\frac{g}{|\Lambda|}a^*_0a_0\sum_{\textbf{k}\neq0}a^*_\textbf{k}a_\textbf{k}\nonumber
\end{equation}
\begin{equation}
+\frac{g}{2|\Lambda|}\sum_{\textbf{k}\neq0}[a^*_0a_0(a^*_\textbf{k}a_\textbf{k}+a^*_{-\textbf{k}}a_{-\textbf{k}})+
a^*_\textbf{k}a^*_{-\textbf{k}}a^2_0+a^{*2}_0a_{-\textbf{k}}a_\textbf{k}]+\frac{g}{2|\Lambda|}a_0^{*2}a_0^2.
\end{equation}
Owing to the macroscopic occupation of the $\textbf{k}=0$ mode at low temperatures, the operators
$a_0/\sqrt{|\Lambda|}$ and $a_0^*/\sqrt{|\Lambda|}$ in $H_\textrm{r}$ can be replaced by complex numbers \={c}, c, respectively. In this manner, Bogoliubov arrives at the Hamiltonian (see Ref. \cite{AVZ92})
\begin{equation}
H_\textrm{B}=\sum_{\textbf{k}\neq0}(\varepsilon_k-\mu+g|\textrm{c}|^2)a_\textbf{k}^*a_\textbf{k}+\nonumber
\end{equation}
\begin{equation}
+\frac{1}{2}g{\sum_{\textbf{k}\neq0}}^{'}[|\textrm{c}|^2a^*_\textbf{k}a_\textbf{k}+|\textrm{c}|^2
a^*_{-\textbf{k}}a_{-\textbf{k}}+
\textrm{c}^2a^*_\textbf{k}a^*_{-\textbf{k}}+{\textrm{\={c}}}^2a_{-\textbf{k}}a_\textbf{k}]+
\frac{1}{2}g|\textrm{c}|^4|\Lambda|-\mu|\textrm{c}|^2|\Lambda|.
\end{equation}
The gauge transformation $a_\textbf{k}\rightarrow a_\textbf{k}\textrm{e}^{i\varphi}$, with adjusted $\varphi$, maps c$\rightarrow|\textrm{c}|$. Let us denote $|\textrm{c}|^2\rightarrow \xi d$, where according to Bogoliubov \cite{NB47}, $\xi$ is the fraction of all atoms belonging to the momentum condensate.

The canonical transformation
\begin{equation}
a_\textbf{k}=u_\textbf{k}b_\textbf{k}+v_\textbf{k}b^*_{-\textbf{k}},\quad u_\textbf{k}^2-v_\textbf{k}^2=1,
\end{equation}
diagonalizes $H_\textrm{B}$ to the form
\begin{equation}
\tilde{H}_{1\textrm{B}}=\sum_{\textbf{k}\neq0}E_{1k}b^*_\textbf{k}b_\textbf{k}
+\frac{1}{2}{\sum_{\textbf{k}\neq0}}^{'}(E_{1k}-f_{1k})-\mu\xi d |\Lambda|+\frac{1}{2}g\xi^2dN,
\end{equation}
where
\begin{equation}
f_{1k}=\varepsilon_k-\mu+2\xi gd,
\end{equation}
\begin{equation}
E_{1k}=\sqrt{f_{1k}^2-h_k^2},\quad h_k=\xi gd,
\end{equation}
and
\begin{equation}
u_k^2=\frac{1}{2}\Big(\frac{f_{1k}}{E_{1k}}+1\Big),\quad v_k^2=\frac{1}{2}\Big(\frac{f_{1k}}{E_{1k}}-1\Big).
\end{equation}

The theory of BEC (e.g. Refs. \cite{FD99}, \cite{LL79}) shows that the ground state of a free Bose gas becomes macroscopically occupied when
\begin{equation}
\lim\mu_\Lambda=e_0\,\,\textrm{as}\,\, |\Lambda|\rightarrow\infty,
\end{equation}
($\mu_\Lambda$ denoting the solution of the equation
\begin{equation}
|\Lambda|^{-1}\textrm{Tr}\,z\exp(-\beta H_0)(1-z\exp(-\beta H_0))^{-1}=d\nonumber
\end{equation}
and the $e_0$ the lowest eigenvalue of $H_0$). The corresponding condition for $\mu$, in the BEC phase of $H_{1\textrm{B}}$, is thus
\begin{equation}
\inf_\mu\lim_{k\rightarrow0} E_{1k}(\mu)=\lim_{k\rightarrow0} E_{1k}(\mu_0)=0,
\end{equation}
under the restriction,
\begin{equation}
E_{1k}(\mu_0)\geq0,\quad\textrm{for all}\,\,\textbf{k}\in\mathbb{R}^3.
\end{equation}
Eq. (56) reduces to
\begin{equation}
(\mu-\xi gd)(\mu-3\xi gd)=0,
\end{equation}
and only the solution $\mu_0=\xi gd$ satisfies $E_{1k}\geq0$, for all $\textbf{k}$. In the BEC phase, the system may be therefore equivalently described in terms of
\begin{equation}
\tilde{H}_\textrm{B}=\sum_{\textrm{k}\neq0}E_{k}b^*_\textbf{k}b_\textbf{k}
+\frac{1}{2}{\sum_{\textbf{k}\neq0}}^{'}
(E_{k}-f_k)-\frac{1}{2}\xi^2 gdN,
\end{equation}
with $\mu=0$ and
\begin{equation}
f_k(\xi,d,a)=\varepsilon_k+\xi gd,\quad E_{k}(\xi,d,a)
=\sqrt{f_k^2-h_k^2},
\end{equation}
\begin{equation}
u_k^2=\frac{1}{2}\Big(\frac{f_k}{E_{k}}+1\Big),\quad v_k^2=\frac{1}{2}\Big(\frac{f_k}{E_k}-1\Big),
\end{equation}
whereas in the gaseous pase, $\tilde{H}_\textrm{B}=H_0$ (Refs. \cite{NB47}, \cite{KH64}, \cite{AVZ92}).

According to the definition (48),
\begin{equation}
{\sum_{\textbf{k}\neq0}}^{'}(E_k-f_k)=\nonumber
\end{equation}
\begin{equation}
=\frac{\hbar^2}{2m}\sum_{\textbf{k}\neq0}[k\sqrt{k^2+2k_a^2}-k^2-k_a^2+\frac{k_a^4}{2k^2}],
\end{equation}
where $k_a^2=8\pi a\xi d$. Passing to the limit $|\Lambda|\rightarrow\infty$, one obtains \cite{LHY57}
\begin{equation}
\frac{1}{2}{\sum_{\textbf{k}\neq0}}^{'}(E_k-f_k)=\frac{\hbar^2|\Lambda|k_a^5}{8\pi^2m}\int_0^\infty dyy^2(y\sqrt{y^2+2}
-y^2-1+\frac{1}{2y^2})=\frac{\sqrt{2}\hbar^2|\Lambda|k_a^5}{15\pi^2m},
\end{equation}
which yields
\begin{equation}
\tilde{H}_\textrm{B}=N\tilde{E}_0(\xi,d,a)
+\frac{\hbar^2}{2m}\sum_{\textbf{k}\neq0}k\sqrt{k^2+2k_a^2}b^*_\textbf{k}b_\textbf{k},
\end{equation}
with
\begin{equation}
\tilde{E}_0(\xi,d,a)=2\pi ad\frac{\hbar^2}{m}[\frac{128}{15}\sqrt{\frac{a^3\xi^5d}{\pi}}-\xi^2].
\end{equation}

Lee and Yang in Ref. \cite{LY58} emphasize the correct linear dependence on $k$ of the excitation spectrum of $\tilde{H}_\textrm{B}$, as $k\rightarrow0$, consistent with the presence of low-energy phonon excitations in liquid $^4{\textrm{He}}$.
As for $\mu$, they do not apply Eq. (56) in the range $T\leq T_\textrm{c}$
in Ref. \cite {LY58} and, as a consequence, the scalar term of the transformed HSBG Hamiltonian is different in their theory.
\begin{figure}
\scalebox{1.1}{\includegraphics{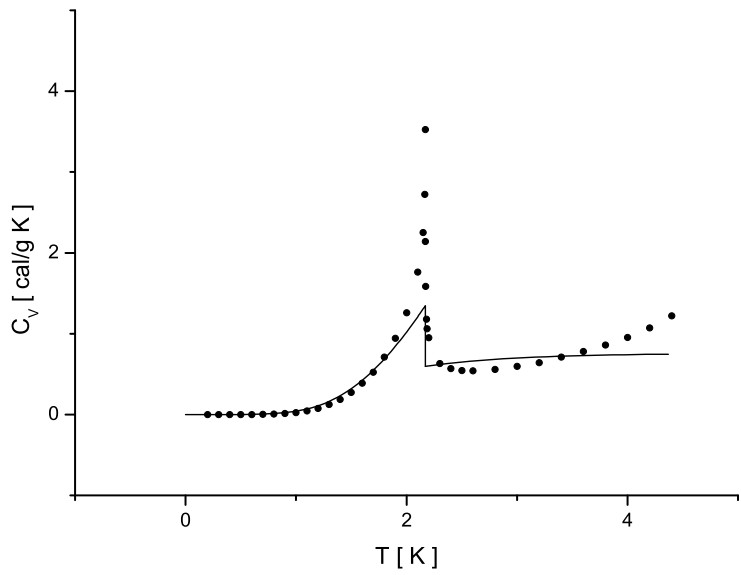}}
\caption{The heat capacity of HSET (Eq. (72)) with $m=m_\textrm{He},\,\,d=d_\textrm{He}=0.02197\,{\AA}^{-3}$ (i.e., under SVP at $T_\lambda$), $\gamma=2.2276\, 10^{-4}\,\textrm{eV},\,a=3\,{\AA}$. The points are experimental results, for varying $^{4}\textrm{He}$ density at SVP, from Ref. \cite{JW63}.}\label{Fi:CVBOGHUA}
\end{figure}

The approach to HSBG thermodynamics, developed in Refs. \cite{LHY57} - \cite{KH95}, which consists in minimizing the free energy
\begin{equation}
\tilde{F}_{1\textrm{B}}(\xi,\beta)=-{\beta}^{-1}\ln\,\textrm{Tr}\exp[-\beta\tilde{H}_{1\textrm{B}}],\nonumber
\end{equation}
(constrained by the equation for $z=\exp(\beta\mu)$) with respect to $\xi$, provides only qualitative agreement of the resulting heat capacity with experimental data on $^{4}\textrm{He}$.

In the present formulation, the free energy of $N$ atoms in the BEC phase equals
\begin{equation}
F_\textrm{B}(\tilde{\beta},\xi,d,a)=NE_0(\xi,d,a)+\frac{4N}{\sqrt{\pi}\tilde{\lambda}^3\tilde{\beta}d}
\int_0^\infty dp\,p^2\ln(1-\exp[-\tilde{E}_p(\xi,d,a)]),
\end{equation}
where
\begin{equation}
\tilde{\lambda}^2=\frac{2\pi\hbar^2\tilde{\beta}}{m},\quad
\tilde{E}_p(\xi,d,a)=\sqrt{p^4+4p^2\tilde{\lambda}^2\xi ad}.
\end{equation}
In this preliminary investigation, we adopt the method presented in Ref. \cite{KH64} (Chapter 3), with the following approximation $\tilde{F}_\textrm{B}(\tilde{\beta},\tilde{\xi}_0,d,a)$ to the free energy:
\begin{equation}
\tilde{F}_\textrm{B}(\tilde{\beta},\tilde{\xi}_0(\tilde{\beta}),d,a):=
F_\textrm{B}(\tilde{\beta},\tilde{\xi}_0(\tilde{\beta}),d,a).
\end{equation}

\textbf{A. The energy and heat capacity}

 The energy $U$ of one gram of $^4{\textrm{He}}$ (expressed in calories) then equals,
\begin{equation}
U(\tilde{\beta},\tilde{\xi}_0,d,a)=5.757230496\,10^3\frac{\partial}{\partial\tilde{\beta}}
\tilde{\beta}\frac{\tilde{F}_\textrm{B}(\tilde{\beta},\tilde{\xi}_0,d,a)}{N}
\end{equation}
(for $\tilde{\beta}$ and $\tilde{F}_\textrm{B}$ expressed in eV), where
\begin{equation}
\frac{\partial\tilde{\beta}\tilde{F}_\textrm{B}}{\partial\tilde{\beta}}=
NE_0(\tilde{\xi}_0,d,a)-\frac{6N}{\sqrt{\pi}\tilde{\lambda}^3\tilde{\beta}d}
\int_0^\infty dp\,p^2\ln(1-\exp(-\tilde{E}_p))\nonumber
\end{equation}
\begin{equation}
+\frac{8Na\tilde{\xi}_0}{\sqrt{\pi}\tilde{\lambda}\tilde{\beta}}
\int_0^\infty \frac{dp\,p^3}{\sqrt{p^2+4\tilde{\lambda}^2a\tilde{\xi}_0d}\,(\exp(\tilde{E}_p)-1)}.
\end{equation}

 The heat capacity of this sample obtains by numerical differentiation of $U$ w.r.t. $T$:
\begin{equation}
C_V(T)=\frac{dU(\tilde{\beta},\tilde{\xi}_0(\tilde{\beta}),d,a)}{dT}.
\end{equation}
Obviously, in $\tilde{F}_\textrm{B}$ and $U_\textrm{B}$ we substitute $d=d_\textrm{He}$ and
\begin{equation}
\tilde{\xi}_0(\tilde{\beta})=
1-\Big(\frac{\tilde{\beta}(\gamma_0,T_\lambda)}{\tilde{\beta}(\gamma_0,T)}\Big)^{3/2},
\quad \textrm{for}\,\,T\leq T_\lambda.
\end{equation}
For $T\geq\tilde{T}_\textrm{c}=T_\lambda,\,\,\xi=0$, hence, $\tilde{H}_\textrm {B}=H_0$ \cite{AVZ92}. In the gaseous phase the energy is thus  given by Eq. (33). By restricting the range of $a$ to values consistent with the graph of the interaction potential $V_\textrm{He}$ \cite {RA79}, we find the resulting $C_V(T)$ plot to be best fitting for $a=a_0=3\,{\AA}$ to experimental data on $^{4}\textrm{He}$ heat capacity measured at SVP. It is shown in Fig. 2. $C_V(T)$ exhibits a peak at $T_\lambda$, which is much smaller than the one observed experimentally, but below 1.8 K, $C_V(T)$ agrees with experiment up to a small error. A comparison of $C_V(T)$ with $^{4}\textrm{He}$ heat capacity, measured at constant density $d_\textrm{He}=0.0225825\,{\AA}^{-3}$ (Ref. \cite{JW63}) reveals similar deviation from experiment. Another point worth emphasizing is the appearance  of singular integrals in the analytic expression for $C_V(T)$, which result by differentiating  the last term on the r.h.s. of Eq. (71) w.r.t. $T$. In the limit $T\rightarrow T_\lambda$, the resulting expression contains a linear combination (with coefficients linear in $\tilde{\xi}_0$ ) of two divergent integrals, viz., $\lim I(x)$ and $\lim K(x)$, as $x\rightarrow0$, where
\begin{equation}
I(x)=\int_0^\infty\frac{dp\,p^3}{(\exp(p\sqrt{p^2+x})-1)(p^2+x)^{3/2}},\nonumber
\end{equation}
\begin{equation}
K(x)=\int_0^\infty\frac{dp\,p^4\exp(p\sqrt{p^2+x})}{(\exp(p\sqrt{p^2+x})-1)^2(p^2+x)}.\nonumber
\end{equation}
We find that
\begin{equation}
\lim_{x\rightarrow 0}\frac{4\sqrt{x}\,I(x)}{\pi}=\lim_{x\rightarrow0}\frac{4\sqrt{x}\,K(x)}{\pi}=1.
\end{equation}
Thus,
\begin{equation}
\lim_{T\rightarrow T_\lambda^{-}}\tilde{\xi}_0(\tilde{\beta})(I(\tilde{\xi}_0(\tilde{\beta}))+
K(\tilde{\xi}_0(\tilde{\beta})))=0,
\end{equation}
so $C_V(T)$ does not diverge as $T\rightarrow T_\lambda^{-}$.
It is well known that the heat capacity of $^4{\textrm{He}}$, measured at constant volume, is finite at $T_\lambda$ \cite{GA73}, but the experimental plot is best modelled by a function singular at $T_\lambda$ \cite{JW63}, \cite{GA73}.
The $C_V(T)$ plot, which obtains by applying the Angelescu-Verbuere-Zagrebnov refinement of Bogoliubov's approximation \cite{AVZ92}, is almost exactly the same as the one in Fig. 2.

\textbf{B. Momentum distribution, momentum condensate and normal fluid density}

Theoretical plots of the momentum distribution (MD), fraction of momentum condensate (FMC) and normal fluid density (NFD) agree with experiment only qualitatively. The plots of normalized MD:
\begin{equation}
\textrm{n}(p,\tilde{\xi}_0(T))=\frac{\textrm{Tr}\Big(\frac{a^*_\textbf{p}a_\textbf{p}}
{\tilde{z}^{-1}\exp[\tilde{\beta}E_p(\tilde{\xi}_0,d,a)b_\textbf{p}^*b_\textbf{p}]-1}\Big)}
{\int d^3k \textrm{Tr}\Big(\frac{a^*_\textbf{k}a_\textbf{k}}{\tilde{z}^{-1}
\exp[\tilde{\beta}E_k(\tilde{\xi}_0,d,a)b_\textbf{k}^*b_\textbf{k}]-1}\Big)},
\end{equation}
for $T=1\,\textrm{K},\,\,2.27\,\textrm{K}$ are shown in Figs. 3 and 4.
\begin{figure}
\scalebox{0.40}{\includegraphics{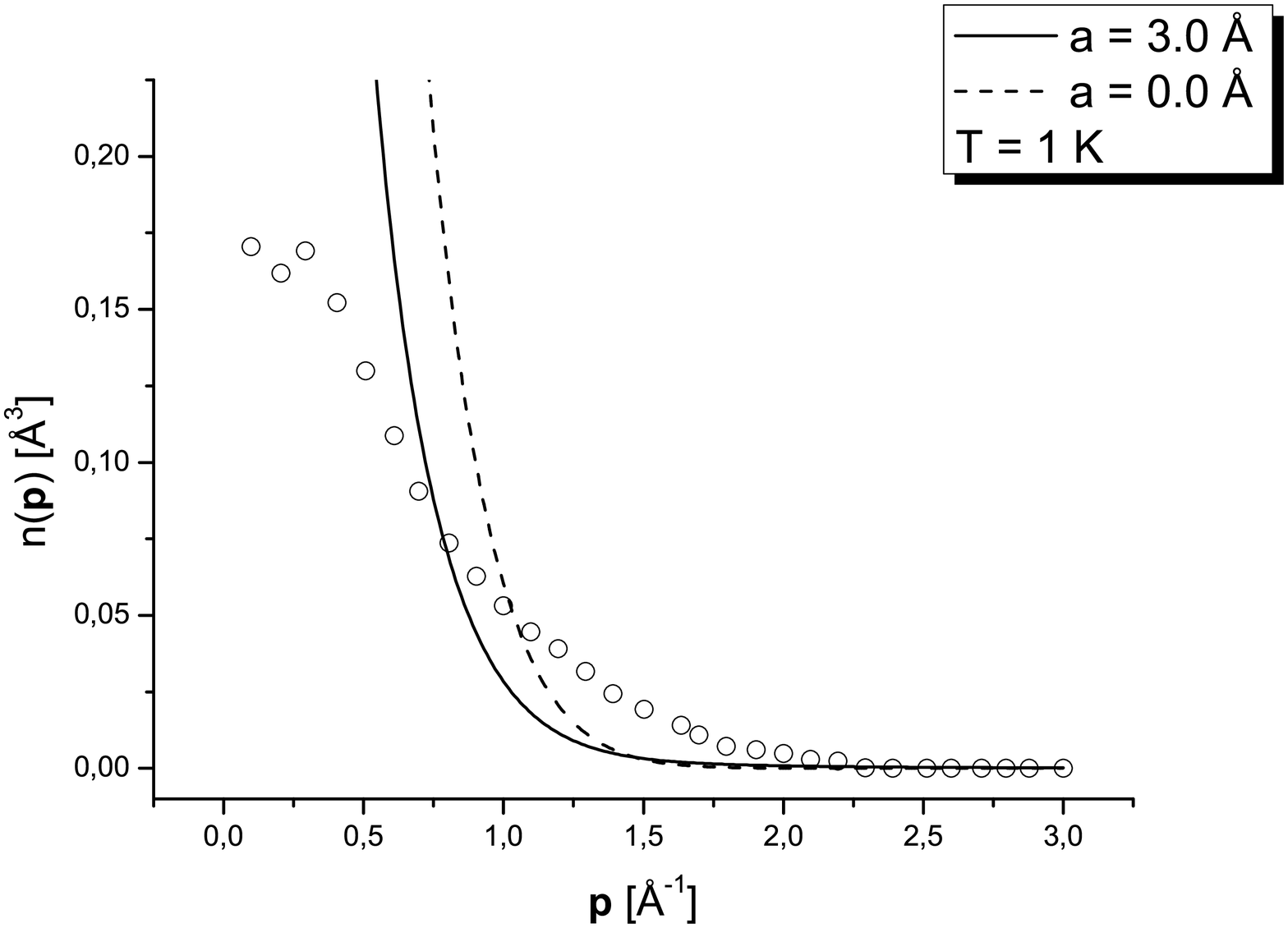}}
\caption{The plots of the momentum distribution function $n(p,\tilde{\xi}_0(T))$ for the FBET and HSET with $a=3\,{\AA}$ at $T=1\,\textrm{K}$. The points are experimental results from Ref. \cite{SSMW82}.}\label{Fi:T1_a3}
\end{figure}
\begin{figure}
\scalebox{0.40}{\includegraphics{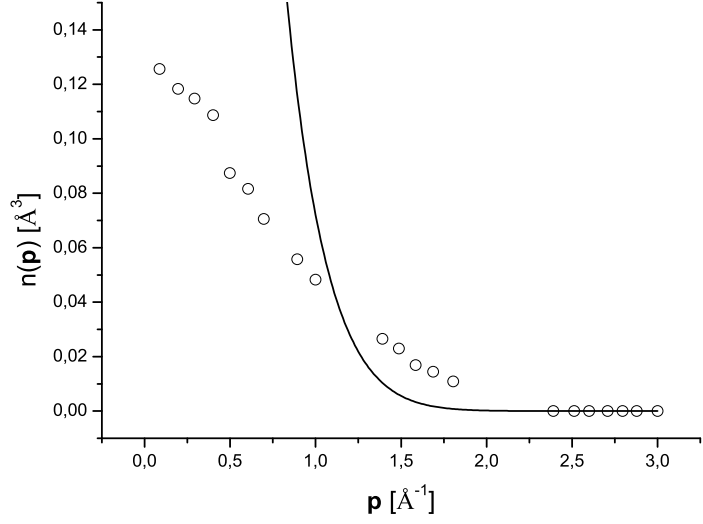}}
\caption{The plot of the momentum distribution function $n(p,0)$ for the FBET and HSET with $a=3\,{\AA}$ at $T=2.27\,\textrm{K}$. The points are experimental results from Ref. \cite{SSMW82}.}\label{Fi:T2_27}
\end{figure}
The FMC, equal
\begin{equation}
\xi(T,\tilde{\xi}_0(T))=1-\frac{1}{N}\sum_{\textbf{k}\neq0}\textrm{Tr}\frac{a_\textbf{k}^*a_\textbf{k}}
{\exp[\tilde{\beta}E_k(\tilde{\xi}_0,d,a) b^*_\textbf{k}b_\textbf{k}]-1}=\nonumber
\end{equation}
\begin{equation}
=1-\frac{2}{\sqrt{\pi}d\tilde{\lambda}^3}\int_0^\infty dpp^2\Big\{\frac{\tilde{f}_p(\tilde{\xi}_0,d,a)}{\tilde{E}_p(\tilde{\xi}_0,d,a)}
\coth\frac{\tilde{E}_p(\tilde{\xi}_0,d,a)}{2}-1\Big\},
\end{equation}
\begin{figure}
\scalebox{0.80}{\includegraphics{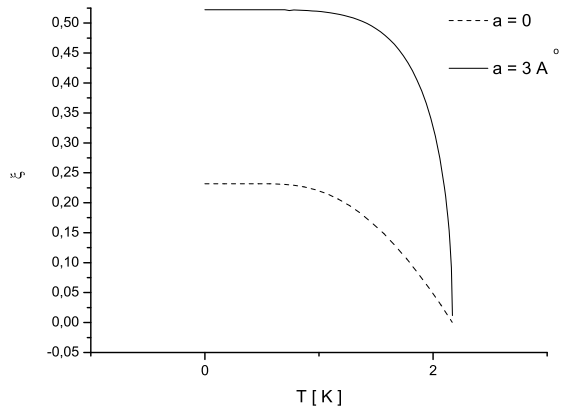}}
\caption{The FMC  $\xi(T,\tilde{\xi}_0(T))$ of FBET ($a=0$) and HSET for $a=3\,{\AA}$. For $^4{\textrm{He}}$ at $T=0\,\textrm{K}$ this fraction is estimated to be between 0.07 and 0.09 \cite{GAS2000}.}\label{Fi:conBOGHUA}
\end{figure}
where $\tilde{f}_p(\tilde{\xi}_0,d,a)=p^2+2a\tilde{\xi}_0\tilde{\lambda}^2d$, is plotted in Fig. 5 for $\gamma=\gamma_0$, $a=a_0$, together with $\tilde{\xi}_0(T)$ for the same $\gamma$. Fig. 6 shows that inclusion of the hard-sphere interaction and treatment of this interaction by the Bogoliubov method, spoils the FMC graph: $\xi(0)=0.5224$, whereas $\tilde{\xi}_0(0)=0.2319$. However, the overall description of HeII is improved (compared to FBET) if, apart from heat capacity, one also takes into account $d_n(T)/d$, where $d_n(T)$ is the normal fluid density (NDF). In HeII, $d_n(0)=0$ and $\xi(0)+d_n(0)/d\approx0.09$. $d_n$ for an interacting Bose gas, in the Bogoliubov approximation, was computed in Ref. \cite{AF70}. The result is analogous to NFD for the FBG:
\begin{equation}
d_n(T,\tilde{\xi}_0(T))=\frac{8}{3\sqrt{\pi}\tilde{\lambda}^3}\int_0^\infty dpp^4\{\textrm{e}^{\tilde{E}_p(\tilde{\xi}_0,d,a)}-1)^{-1}+
(\textrm{e}^{\tilde{E}_p(\tilde{\xi}_0,d,a)}-1)^{-2}\}.
\end{equation}
$d_n(T,\tilde{\xi}_0(T))/d$ is plotted in Fig. 7. One finds, $d_n(0,\tilde{\xi}_0(0))/d=0.1896$ for $\gamma=\gamma_0,\,\,a=a_0$. Thus
\begin{equation}
\xi(0)+\frac{d_n(0,\tilde{\xi}_0(0))}{d}=0.712,
\end{equation}
which is an improvement, as regards $^{4}\textrm{He}$ thermodynamics, of the FBG result
\begin{equation}
\xi_0(0)+\frac{d_{0n}(0)}{d}=1.
\end{equation}
The plot of $d_n(T,\tilde{\xi}_0(T))/d$, for $^4{\textrm{He}}$ values of $d$, $m$ and $a= 3\,{\AA}$, is depicted in Fig. 6.
\begin{figure}
\scalebox{0.8}{\includegraphics{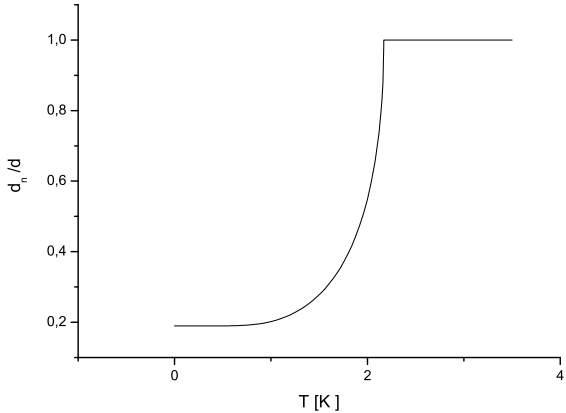}}
\caption{The fraction of HSET normal fluid (Eq. (78)) for $d=d_\textrm{He},\,\,m=m_\textrm{He},\,\,a=3\,{\AA}$. In $^4{\textrm{He}}$ at $T=0\,\textrm{K}$ this fraction  equals zero.}\label{Fi:dnBOGHUA}
\end{figure}

\textbf{C. Latent heat of the transition}

The pressure of the HSET in the BEC phase, as implied by Eq. (67), equals
\begin{equation}
P=-\frac{4}{\sqrt{\pi}\tilde{\beta}\tilde{\lambda}^3}\int_0^\infty\,dq\,q^2\ln(1-\exp[-\tilde{E}_q])
-dE_0(\tilde{\xi}_0,d,a).
\end{equation}
The Clapeyron equation (41), for $P$ given by Eq. (81), yields the same latent heat of the BEC transition for the HSET as for the FBET.

\section{Hard-sphere Bose gas with dressed condensate bosons}

Creation and annihilation processes of atoms in the momentum condensate of a Bose gas are represented by the operators $a^*_0,\,\,a_0$, which appear, in the Bogoliubov-Huang theory, on the same footing as those describing atoms with nonzero momenta. However, owing to their immobility and the small fraction they constitute (e.g. in helium II), the condensate atoms can be expected to be particularly predisposed to attract the excited ones and thus act as centres surrounded by uncondensed atoms. From such point of view, it would be more appropriate to treat the condensate bosons as atoms surrounded by excited bosons. This can be achieved by performing a unitary transformation $a_0\rightarrow\alpha_0$ in $H$, with $\alpha_0,\,\alpha^*_0$ representing the dressed bosons in the condensate. This type of procedure was applied by Fr\"{o}hlich in Ref. \cite{HF52} to the electron-phonon Hamiltonian $H_{\textrm{el-ph}}$. The physically meaningful terms of the transformed $H_{\textrm{el-ph}}$ were incorporated by Bardeen, Cooper and Schrieffer into their theory of superconductivity \cite{BCS57}.

We consider the simple displacement transformation
\begin{equation}
\alpha_0=a_0+\nu,
\end{equation}
with $\nu\epsilon\mathbb{R}^1$ representing the cloud of excited bosons accompanying the immobile one, as a possible trial realization of the mapping $a_0\rightarrow\alpha_0$. Glauber demonstrated in Ref. \cite{RG63} that the mapping (82) can be accomplished by the unitary transformation
 \begin{equation}
 D(\nu)^{-1}a_0D(\nu)=a_0+\nu,
  \end{equation}
 where
 \begin{equation}
 D(\nu)=\exp[\nu(a_0^*-a_0)].
 \end{equation}

In the transformed Hamiltonian $H_{1D}=D(\nu)^{-1}H_\textrm{r}D(\nu)$, we perform Bogoliubov's substitution $a_0\rightarrow\sqrt{\xi d|\Lambda|}$ and, furthermore, $\nu\rightarrow\sqrt{\eta d|\Lambda|}/2$, where $\eta/4>0$ is the fraction of bosons which participate in the dressing of those in the momentum condensate. Since $0<\xi\ll1$ in $^4{\textrm{He}}$, therefore $\eta/4$ can be also expected to be small. Thus $\sqrt{\xi}>\xi,\,\,\sqrt{\eta}/4>\eta/4$, so the terms of the reduced interaction in $H_{1D}$, most significant for the description of BEC in $^4{\textrm{He}}$, are those linear in $\sqrt{\xi\eta}$ and $\xi$. The latter were considered in Sec. 4. In this section, we take into account only the former ones and disregard all others. In this manner, we obtain the Hamiltonian
\begin{equation}
H_D=\sum_{\textbf{k}\neq0}(\varepsilon_k-\mu+g\sqrt{\xi\eta}\,d)a^*_{\textbf{k}}a_\textbf{k}+\nonumber
\end{equation}
\begin{equation}
\frac{1}{2}g\sqrt{\xi\eta}\,d{\sum_{\textbf{k}\neq0}}^{'}(a^*_\textbf{k}a_\textbf{k}+a^*_{-\textbf{k}}a_{-\textbf{k}}+
a^*_\textbf{k}a^*_{-\textbf{k}}+a_{-\textbf{k}}a_\textbf{k})-\mu\sqrt{\xi\eta}\,d|\Lambda|.
\end{equation}

It should be noted at this point that Bogoliubov rejected in his paper \cite{NB47} all terms of the 2-body interaction linear in $a_0,\,\,a^*_0$, such as
\begin{equation}
\frac{g}{2|\Lambda|}
\sum_{\begin{subarray}{l}
\textbf{p},\,\textbf{q}\neq0\\
\textbf{p}-\textbf{q}\neq0
\end{subarray}}a_0^*a^*_\textbf{q}a_\textbf{p}a_{\textbf{q}-\textbf{q}},\quad
\frac{g}{2|\Lambda|}\sum_{\begin{subarray}{l}
\textbf{p},\,\textbf{q}\neq0\\
\textbf{p}-\textbf{q}\neq0
\end{subarray}}a^*_{\textbf{p}-\textbf{q}}a^*_\textbf{q}a_\textbf{p}a_0.
\end{equation}
 These terms cannot be included into Bogolibov's diagonalization procedure. Processes described by the expressions (86), representing the creation or annihilation of a single boson in the momentum condensate, as a result of 2-body interactions, should be expected to occur especially at temperatures $T$ satisfying $0<1-T/T_\lambda\ll1$, where the number of such atoms is macroscopic, but still very small. $H_D$ can be viewed as a Hamiltonian which simulates such processes.

 The transformation (51) diagonalizes $H_D$ and the same rearrangements , as applied in Sec. 4 to $H_B$, transform $H_D$ to the form
 \begin{equation}
 \tilde{H}_D=N\tilde{E}_D(\xi,d,a_D)+
 \frac{\hbar^2}{2m}\sum_{\textbf{p}\neq0}p\sqrt{p^2+2k_D^2}\,b^*_\textbf{p}b_\textbf{p},
 \end{equation}
 where
 \begin{equation}
 \tilde{E}_D(\xi,d,a_D)=2\pi a_Dd\frac{\hbar^2}{m}\Big(\frac{128}{15}\sqrt{\frac{a_D^3\xi^{5/2}d}{\pi}}-2\xi\sqrt{\eta}\Big),
 \quad k_D^2=8\pi\sqrt{\xi}\,a_Dd,\quad a_D=a\sqrt{\eta}.
 \end{equation}

 In the sequel we put $\eta=1$, in order to restore the unscaled hard-sphere diameter $a$. The system described by $H_D$, with $\eta=1$, will be called the $D$-transformed HSET (DHSET).

 \textbf{A. The energy and heat capacity}

 For $\eta=1$, the energy of 1 gram of $^4{\textrm{He}}$ (expressed in calories) under the same approximation $\xi=\tilde{\xi}_0$, as in Sec. 4, equals
 \begin{equation}
 U(\tilde{\beta},\tilde{\xi}_0^{1/2},d,a)=5.757230496\,10^3\big(\tilde{E}_0(\tilde{\xi}_0^{1/2},d,a)
 -\frac{6N}{\sqrt{\pi}\tilde{\lambda}^3\tilde{\beta}d}\int_0^\infty dp\,p^2\ln(1-\exp(-\tilde{E}_p(\tilde{\xi}_0^{1/2},d,a)))\nonumber
 \end{equation}
 \begin{equation}
 +\frac{8Na\tilde{\xi}_0^{1/2}}{\sqrt{\pi}\tilde{\lambda}\tilde{\beta}}\int_0^\infty\frac{dp\,p^3}
 {\sqrt{p^2+4a\tilde{\lambda}^2\tilde{\xi}_0^{1/2}\,d}\,(\exp(\tilde{E}_p(\tilde{\xi}_0^{1/2},d,a))-1)}\big)
 \end{equation}
 (for $\tilde{\beta},\,\,\tilde{\lambda}^2$ expressed in eV). The square root $\tilde{\xi}_0(\tilde{\beta})^{1/2}$ appearing in the coefficient of the second integral on the r.h.s. leads to the singularity
 \begin{equation}
 C_V(T)\approx\frac{\textrm{const}}{(T_\lambda-T)^{1/2}}
 \end{equation}
 of heat capacity, as $T\rightarrow T_\lambda^{-}$. The best fitting to experiment heat capacity graph, obtained for $d=d_{\textrm{He}}, m=m_{\textrm{He}}$ and $\gamma=\gamma_0,\,\,a=2.45{\AA},\,\,$, by differentiating $U(\tilde{\beta},\tilde{\xi}_0(\tilde{\beta})^{1/2},d,a)$ numerically w.r.t. $T$,
 \begin{equation}
 C_V(T)=\frac{U(\tilde{\beta},\tilde{\xi}_0(\tilde{\beta})^{1/2},d,a)}{dT},
 \end{equation}
 is depicted in Fig. 7. There is good quantitative agreement with experimental data on $^4{\textrm{He}}$ heat capacity at SVP below 2.1 K. For $T\in[2.1\,\textrm{K},\,2.17\,\textrm{K}]$ the $C_V$, given by Eq. (91), exceeds the experimental values. It is therefore indeed the temperature range $0<1-T/T_\lambda\ll1$, where the $C_V(T)$ plot in Fig. 2 is modified by the Hamiltonian $\tilde{H}_D$ most significantly. Furthermore, the best fitting value $a=$ 2.45 {\AA} for Eq. (91), agrees with smaller error, than $a_0$ in Sec. 4, with the empirically established hard-sphere diameter $a\approx$ 2.6 {\AA} of $^4{\textrm{He}}$ atoms \cite{RA79}.

 The latent heat of the $\tilde{H}_D$ BEC transition is the same as for $\tilde{H}_B$.
 \begin{figure}
 \scalebox{1.1}{\includegraphics{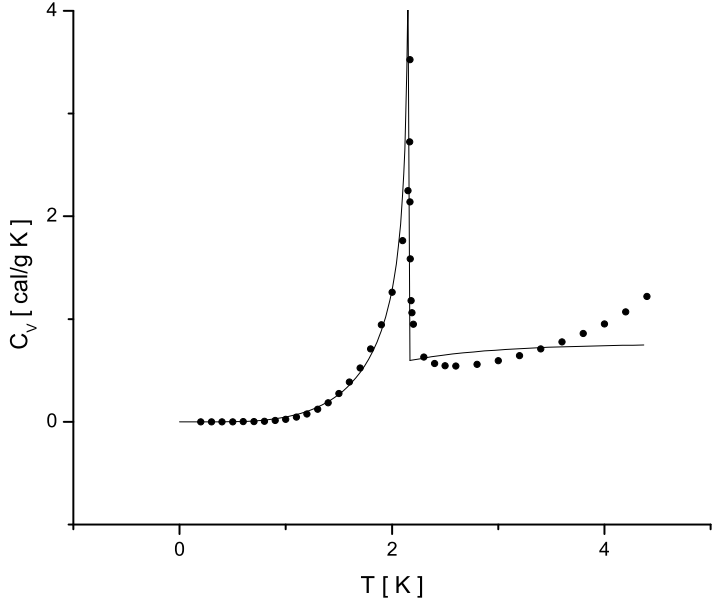}}
 \caption{The heat capacity $C_V(T)$, given by Eq. (91), for $d=d_{\textrm{He}},\,\,m=m_{\textrm{He}},\,\,a=2.45\,{\AA},\,\,\gamma=\gamma_0$. The points are experimental heat capacity results from Ref. \cite{JW63}. }\label{Fi:CVDRESSED}
 \end{figure}
\newpage
 \textbf{B. Momentum distribution, momentum condensate and normal fluid density}

 The DHSET Hamiltonian improves the theoretical plots of MD, FMC and NFD obtained in Sec. 4. These quantities for the DHSET, equal $n(p,\tilde{\xi}_0(T)^{1/2}),\,\,\xi(T,\tilde{\xi}_0(T)^{1/2}),\,\,d_n(T,\tilde{\xi}_0(T)^{1/2})$, respectively, are plotted in Figs. 8, 9, 10. For $T=2.27$ K, the plot of $n(p,\tilde{\xi}_0(2.27\,\textrm{K})^{1/2})$ is the same as in Fig. 4.
 \begin{figure}
 \scalebox{.50}{\includegraphics{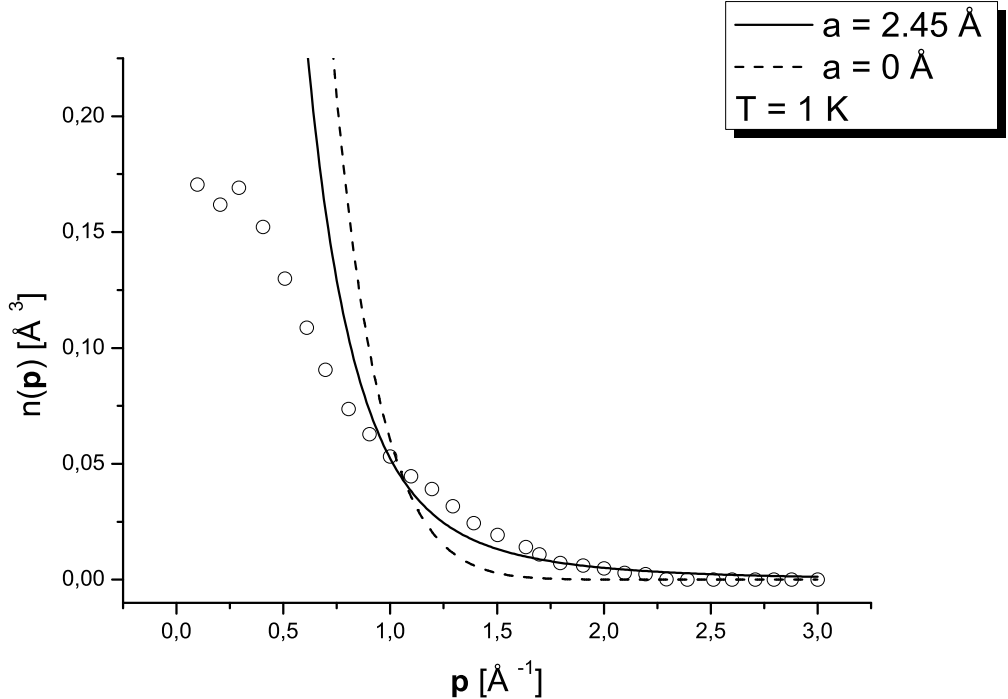}}
 \caption{The plots of the MD function $n(p,\tilde{\xi}_0(1)^{1/2})$ for the FBET and DHSET with $a=2.45$ {\AA}, $\gamma=\gamma_0$ at $T=1$ K. The points are experimental results from Ref. \cite{SSMW82}}.\label{Fi:T1}
 \end{figure}
 \begin{figure}
 \scalebox{1}{\includegraphics{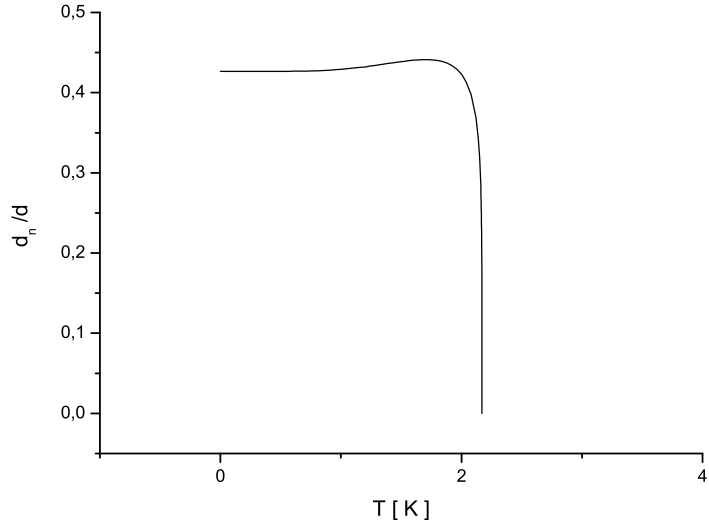}}
 \caption{The FMC function $\xi(T,\tilde{\xi}_0(T)^{1/2})$ of DHSET for $a=2.45$ {\AA} $\gamma=\gamma_0$, $d=d_{\textrm{He}}, \,\, m=m_{\textrm{He}}$. For $^4{\textrm{He}}$ at $T=0$ K this fraction is estimated to be between 0.07 and 0.09. \cite{GAS2000}}\label{Fi:CONDRES}
 \end{figure}
 \begin{figure}
 \scalebox{1}{\includegraphics{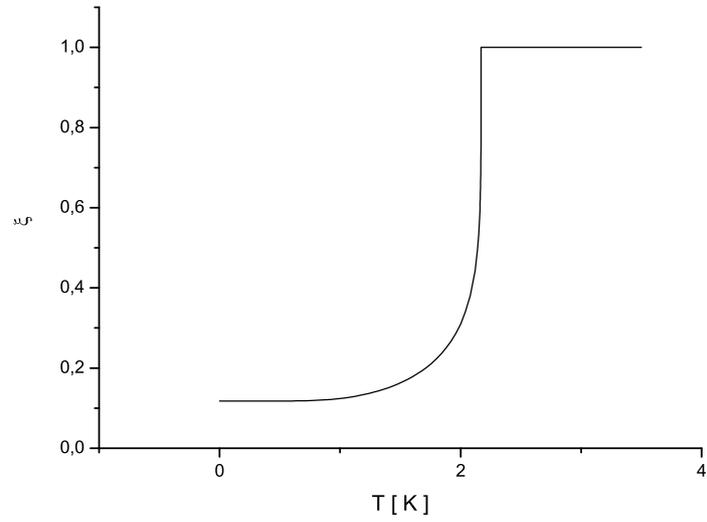}}
 \caption{The fraction of DHSET normal fluid density $d_n(T,\tilde{\xi}_0(T)^{1/2})/d$ for $a=2.45{\AA},\,\,\gamma=\gamma_0,\,\,d=d_{\textrm{He}},\,\,m=m_{\textrm{He}}$. In $^4{\textrm{He}}$ at $T=0$ K this fraction is equal zero.}\label{Fi:GNDRES}
 \end{figure}

 These results, as well as the DHSET heat capacity plot in Fig. 7, show that the DHSET description of helium II thermodynamics considerably improves the results of HSET theory. In particular, $\xi(0,\tilde{\xi}_0(0)^{1/2})=0.4266$, $d_n(0,\tilde{\xi}_0(0)^{1/2})/d=0.1178$ and
 \begin{equation}
 \xi(0,\tilde{\xi}_0(0)^{1/2})+\frac{d_n(0,\tilde{\xi}_0(0)^{1/2})}{d}=0.5444,
 \end{equation}
 which is an improvement of the corresponding HSET result (Eq. (79)) by 23.5\%.

\section{Concluding remarks}

The DHSET has revealed good quantitative agreement of heat capacity with experimental $^{4}\textrm{He}$ measurements below 2.1 K and considerable improvement of the HSET heat capacity graph, although the sharp, but continuous, decrease of heat capacity observed above $T_\lambda$ in $^{4}\textrm{He}$, is not reproduced the HSET nor DHSET. Other thermal properties of the DHSET also improve the HSET results, but agree with experimental data on $^{4}\textrm{He}$ only qualitatively. Further improvement of this theory is therefore necessary. One possible extension could include a more general form of effective temperature, another inclusion of all terms of $H_{1D}$, linear both in $\sqrt{\xi\eta}$ and $\xi$, into the theory.


\begin{thebibliography}{99}
\bibitem{PS91}
P. E. Sokol, W. M. Snow in \emph{Excitations in 2D and 3D Quantum Fluids}, ed. by A. F. G. Wyatt and H. J. Lauter, Plenum, New York, 1991
\bibitem{FL38}
F. London, Nature 141, 643 (1938)
\bibitem {NB47}
N. N. Bogoliubov, J. Phys. (USSR) 11, 23 (1947)
\bibitem{NB70}
N. N. Bogoliubov, Lectures on Quantum Statistical Mechanics in \emph{Quantum Statistics, Vol. 1}, Gordon and Breach, 1970
\bibitem{LHY57}
T. D. Lee, K. Huang, C. N. Yang, Phys. Rev. 106, 1135 (1957)
\bibitem{LY58}
T. D. Lee, C. N. Yang, Phys. Rev. 112, 1419 (1958)
\bibitem{KH64}
K. Huang in \emph{Studies in Statistical Mechanics, Vol. II}, ed. by J. de Boer and G. E. Uhlenbeck, North Holland, Amsterdam, 1964
\bibitem{KH95}
K. Huang in \emph{Bose-Einstein Condensation}, ed. by A. Griffin, D. W. Snoke and S. Stringari, Cambridge University Press, 1995
\bibitem{AVZ92}
N. Angelescu, A. Verbeure, V. A. Zagrebnov, J. Phys. A : Math. Gen. 25, 3473 (1992)\emph{}
\bibitem{GAS2000}
H. R. Glyde, R. T. Azuah, W. G. Stirling, Phys. Rev. B 62, 14337 (2000)
\bibitem{AL93}
A. Griffin, \emph{Excitations in a Bose-Condensed Liquid}, Cambridge University Press, 1993
\bibitem{CP86}
D. M. Ceperley, E. L. Pollock, Phys. Rev. Lett. 56, 351 (1986)
\bibitem{PC87}
E. L. Pollock, D. M. Ceperley, Phys. Rev. B 36, 8343 (1987)
\bibitem{DMC95}
D. M. Ceperley, Rev. Mod. Phys. 67, 279 (1995)
\bibitem{RA79}
R. A. Aziz, V. P. S. Nain, J. S. Carley, W. L. Taylor, G. T. McConville, J. Chem. Phys. 70, 4330 (1979)
\bibitem{MZ00}
M. A. Za{\l}uska-Kotur, M. Gajda, A. Or{\l}owski, J. Mostowski, Phys. Rev. A 61, 033613 (2000)
\bibitem{FD99}
F. Dalfovo, S. Giorgini, L. P. Pitaevskii, S. Stringari, Rev. Mod. Phys. 71, 463 (1999)
\bibitem{SP71}
S. Pruski, J. Ma\'{c}kowiak, Rep. Math. Phys. 1, 309 (1971)
\bibitem{SPA72}
S. Pruski, J. Ma\'{c}kowiak, O. Missuno, Rep. Math. Phys. 3, 227 (1972)
\bibitem{SPB72}
S. Pruski, J. Ma\'{c}kowiak, O. Missuno, Rep. Math. Phys. 3, 241 (1972)
\bibitem{SP74}
J. Ma\'{c}kowiak, O. Missuno, S. Pruski, Rep. Math. Phys. 5, 327 (1974)
\bibitem{JM99}
J. Ma\'{c}kowiak, Phys. Rep. 308, 235 (1999)
\bibitem{LL51}
L. D. Landau, E. M. Lifshitz, \emph{Statistical Physics}, Nauka, Moscow, 1951
\bibitem{JM07}
J. Ma\'{c}kowiak, Open Sys. Information Dyn. 14, 229 (2007)
\bibitem{JM08}
J. Ma\'{c}kowiak, Open Sys. Information Dyn. 15, 281 (2008)
\bibitem{KH63}
K. Huang, \emph{Statistical Mechanics}, J. Wiley, Inc., New York, London, 1963
\bibitem{JW63}
J. Wilks, \emph{Properties of Liquid and Solid Helium}, Clarendon, Oxford, 1967
\bibitem{GA73}
G. Ahlers, Phys. Rev. A 8, 530 (1973)
\bibitem{LL79}
L. Landau, I. F. Wilde, Commun. Math. Phys. 70, 43 (1979)
\bibitem{SSMW82}
V. F. Sears, E. C. Svensson, P. Martel, A. D. B. Woods, Phys. Rev. Lett. 49, 279 (1982)
\bibitem{AF70}
A. F. Fetter, Ann. Phys. 60, 464 (1970)
\bibitem{MB10}
J. Ma\'{c}kowiak, D. Borycki, Mod. Phys. Lett. B 24, 2131 (2010)
\bibitem{HF52}
H. Fr\"{o}hlich, Proc. R. Soc. London, Ser. A 215, 291 (1952)
\bibitem{BCS57}
J. Bardeen, L. N. Cooper and J. R. Schrieffer, Phys. Rev. 108, 1175 (1957)
\bibitem{RG63}
R. J. Glauber, Phys. Rev. 131, 2766 (1963)

\end{thebibliography}
\end{document}